\documentstyle[12pt]{article}
\begin{document}

\begin{center}{\bf \Huge Radially Interrupted Viscous Fingers in a Lifting Hele-Shaw Cell}
\vskip .5cm
Suparna Sinha$^1$, Subrata Kumar Kabiraj$^1$, Tapati Dutta$^2$   and Sujata Tarafdar$^1$\\
$^1$ Condensed Matter Physics Research Centre,\\Physics Department, Jadavpur University,\\
Kolkata 700032, India.\\Corresponding author: S Tarafdar (email : sujata@juphys.ernet.in)\\
$^2$ Physics Department, St. Xavier's College,\\
Kolkata 700016, India\\
\end{center}

\noindent {\bf Abstract}\\Viscous fingers have been produced in  the lifting Hele-Shaw cell, with concentric 
circular grooves etched onto the lower plate. The invading fluid (air) enters the defending newtonian fluid - olive oil
as fingers proceeding radially inwards towards the centre. The fingers are interrupted at the circular groove, and reform
as secondary fingers. The effect of the grooves is to speed up the fingering process considerably and the fingers now 
reach the centre  much faster. We explain this by comparing the variation in velocity of the fingers in the normal HS cell
and the grooved cells with time. In the normal HS cell the fingers move fastest on initial formation and slow down later. 
Since in  case of the grooved plate, the fingers reform  and receive a boost in their speed each time they encounter a 
groove, the fingers proceed to the centre faster.\\
PACS nos. 47.20.Gv, 47.54.+r, 68.03.-g
\vskip .5cm
Viscous fingers (VF) are formed when a fluid of lower viscosity is forced into a fluid
of higher viscosity \cite{vf1,rev1}. 
The interface between the two fluids, develops finger-like intrusions which sometimes branch repeatedly
into fractal patterns. Viscous fingering belongs to the well-known family of Laplacian growth phenomena
which include for example - diffusion limited aggregation, dendritic growth and dielectric breakdown \cite{vic}.

The process is studied conveniently  in the Hele-shaw (HS)
cell, where the more viscous defending fluid is confined in a narrow gap between two
glass plates. Characteristics of the pattern depend on viscosities of the fluids, interface tension between them
and also details of the experimental arrangement and forcing pressure.
In the conventional HS cell, the invading fluid enters the gap through a
hole in the centre of the upper plate. In the Lifting Hele-Shaw cell (LHSC), the upper
plate is slowly lifted, either from one end \cite{lhsc1}, or 
keeping it parallel to the lower plate, in this case  the invading fluid enters
from the sides forming a radial pattern \cite{st}. The LHSC has not been studied
as extensively as the conventional HS cell. It is useful for investigation of adhesion
and is receiving a lot of attention at present \cite{bonn,gay}.

The Hele-shaw cell can be modified in a number of ways to produce interesting
variation in the patterns \cite{mm}. One way is to etch grooves onto the lower plate, this
introduces an anisotropy which affects the patterns \cite{mm}, or stretch  a piece of cloth across the lower plate \cite{amar}.
It has also been shown that besides the symmetry of the superposed pattern the local geometry also plays a  crucial role 
\cite{ab}. So far there has however, been no report of study of patterns in the LHSC with any such perturbations.

In the present communication we report a study of LHSC patterns with concentric
circular grooved etched on the lower plate. The etched pattern is radially symmetric
conforming to  the LHSC arrangement and the fingers entering the cell from the sides are
interrupted normally. This causes the fingering pattern to be disrupted and reformed
each time it meets a groove. The appearance and width of the fingers change and the
most surprising result is that the entire fingering pattern proceeds much faster and
the fingers now reach the centre in a fraction of the time it takes without the grooves.

Our lifting Hele-Shaw cell consists of two thick $(\sim 0.5cm)$ glass plates. The lower one is fixed to a rigid frame,
and the upper one can be lifted using a pneumatical cylinder arrangement. The lifting force can be adjusted. In the
 standard LHSC for studying adhesion the velocity of plate separation is constant, whereas in our apparatus the lifting force
is controlled and kept constant during the experiment, allowing the velocity to vary. The process of pattern formation is
recorded by a CCD camera (at the rate of 25 frames per second) and analysed using image-pro plus software. 
The defending fluid is placed on the lower plate, and the upper plate pressed down on it. This makes the fluid form a circular blob with  diameter of several cm., the upper plate is then lifted slowly.

In the present modification we use two different lower plates, one has a single
circular groove (diameter 19 mm) etched on it, and the other has three concentric grooves
(diameter 9 mm, 17.5 mm and 24.5 mm). The width of the grooves is about 0.1 cm, and depth $\sim 0.05  cm$. 
 The patterns are compared with normal LHSC with a plane lower plate. The
defending fluid used here is olive oil (coloured slightly with a dye), this is a
newtonian fluid and the invading fluid is air (assumed to have zero viscosity).

Figure 1(a-b) shows two  successive stages in pattern formation, when the lower plate
has no grooves. Figure 2(a-b) and figure 3(a-b) show respectively the patterns with
one and three grooves. Table I shows the results of measurements - we take a measured volume of 
olive-oil using a micro-pipette.
Different sets are taken with different force of separation, in the present arrangement though the force can be varied, we
cnnot measure it precisely. The diameter of the initial circular pattern before fingering
starts is given, $t_c$ is the time required for the fingers to reach the centre of the pattern and $t_s$, is the
time required for complete separation of the plates. The results show that for the 3-grooved plate, the time for
fingers to reach the centre is always much smaller than for the plane plate. for the single grooved plate, the effect is
not for noticable. The time required for the plates to separate
completely is however, more or less independent of the presence of grooves. This is expected, since the total energy required to separate the plates should depend on the radius of the iniital blob and its thickness only.

Each result shown in the table is an average over 4-7 patterns.
The fingering is a stochastic process and the patterns are not perfectly symmetric, so the fingering times reported
are not exact quantities, but the gross characteristics can be identified in spite of a certain amount of uncertainty.

To explain the difference in $t_c$ between  patterns with and without grooves, 
we look at the finger velocities in the two  cases . Figure 2 shows how the velocity  of the fastest finger  varies with time. In figure 2a we show velocity vs. time for a typical pattern without grooves. The
velocity is seen to be highest at the outset, it later undergoes some fluctuations but
on the whole decreases, ultimately going to zero. In the grooved plate patterns we
observe that as soon as one or two of the competing fingers reach the largest groove
the air spreads very rapidly around the groove, and then start forming secondary
fingers from the groove, which look exactly like the initial fingers. the secondary
fingers again start with the very high initial velocity of the primary fingers. this
process is repeated at the next groove. So on the whole the fingers proceed towards the
centre with a much higher average velocity compared to  the
plane plate, making $t_c$ much smaller. For the single-grooved plate, the lifting force in set-II is much smaller,
so the fingers do not get time to speed up within the groove. we are at present, studying the effect of different lifting forces more thoroughly.

In the grooved plates there is a strong tendency towards cavitation. It is quite difficult to eliminate air-bubbles completely
and  it is possible that minute quantities of air remain trapped in the grooves or in
irregularities in the plates, these are initially too small to be seen by the naked eye, but start expanding as the upper
late is lifted. Cavitation and its role in adhesion has been discussed previously by Gay et al \cite{gay}.
The initial width and velocity of fingers has been studied theoretically by linear stability analysis \cite{sta}, but the 
development of finger velocities upto breakthrough or upto complete separation of the plates in LHSC has not
been studied in detail. Shelley et al \cite{ms} have studied the full formation of the LHSC pattern numerically, their
results indicate that in a newtonian fluid the finger velocity first increases, but then drops off as the fingers
approach the centre of the plate. This is similar to what we observe with the newtonian fluid. However, earlier
work with a non-newtonian fluid showed that the 5-6 fingers which reach the centre, speed up at the end \cite{kabi}.

We are continuing the study of VF patterns with circular grooves - we plan to use different fluids, newtonian and 
non-newtonian and vary the number and spacing of the grooves. we hope to report more detailed results in the near future.\\
\vskip .5cm \noindent
{\bf Acknowledgement}: This work is supported by the "Indo-French Centre
for the Promotion of Advanced Research/ Centre Franco-Indien Pour le Promotion
de la Rechereche Advancee" through a joint  project in collaboration with M Ben Amar and Y Couder, Laboratoire de
Physique Statistique, ENS Paris. Authors thank Minakshi Maitra for helping with the experiments.

 \vskip .5cm

\noindent  {\bf Table I} Fingering in LHSC with the lower plate having no groove (plane), one groove (1-grv) and
three grooves (3-grv). Lifting force is different in the different sets. The initial diameter of the fluid blob (diam),
the time for fingers to reach the centre ($t_c$), and the time of separation ($t_s$) are shown.
\vskip 1cm
\begin{tabular}{|c|c|c|c|c|c|}
\hline 
set        & type  & vol ($\mu$l) & diam. (cm) & $t_c$ (sec) & $t_s$ (sec) \\ \hline

I         & plane &  150        & 5.2       &  4.6        &   7.9 \\
           & 3-grv &  150        & 4.5       &  1.3        &   8.0 \\ \hline
II        & plane &  150        & 5.7       & 28.5        &  34.5 \\
           & 1-grv &  150        & 5.7       & 28.1        &  34.7 \\
           & 3-grv &  150        & 5.5       &  0.9        &  32.9 \\ \hline
\end{tabular}     
\vskip .5 cm \noindent {\bf Figure captions}
\vskip .5 cm \noindent Figure 1. (a) and (b) showan early and final stage of pattens with the plane lower plate, 
(c) and (d) are for the single-grooved plate and (e) and (f) for the three-grooved plate.
\vskip .5 cm \noindent Figure 2. (a) and (b) show velocity of the fastest finger vs. time for the plane and three-grooved plate respectively. In (b) $v_1$ is before reaching the largest groove, $v_2$ between largest and middle groove, $v_3$ between
middle and smallest groove and $v_4$ from smallest groove to centre.

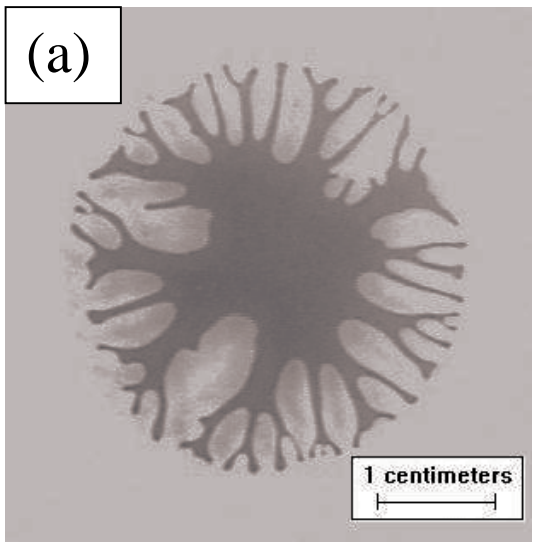
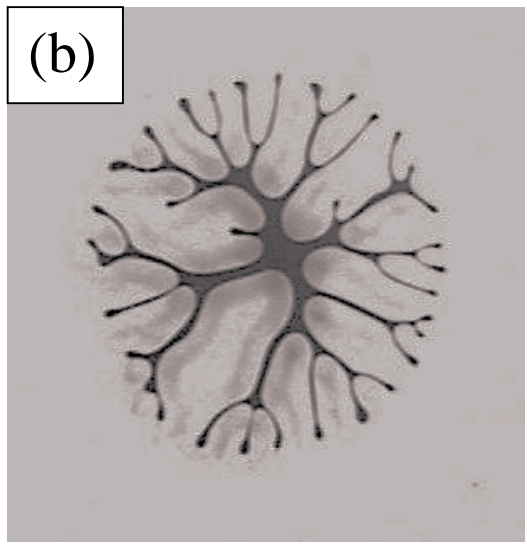
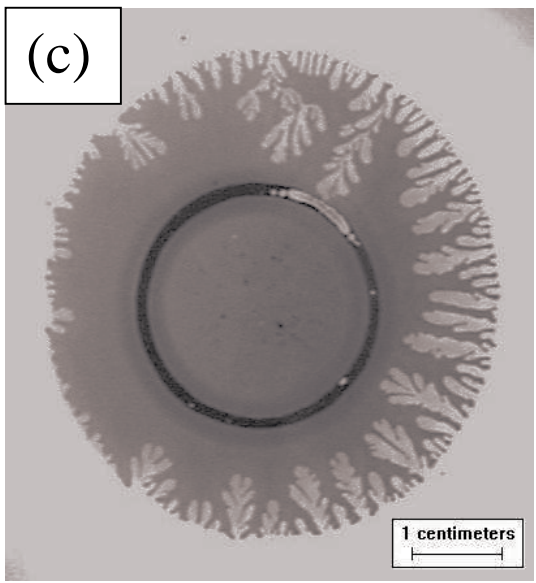
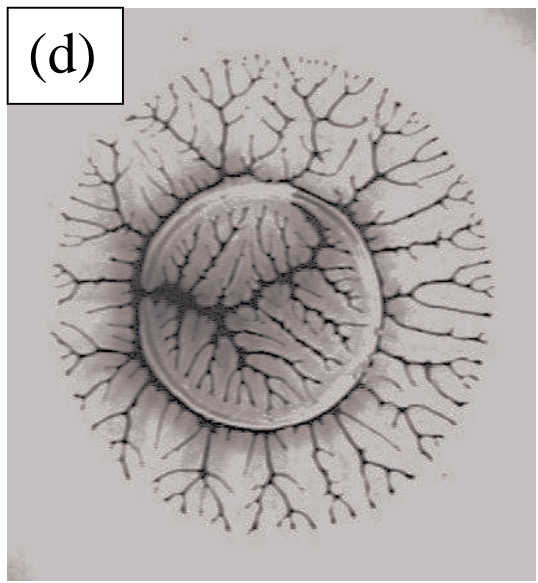
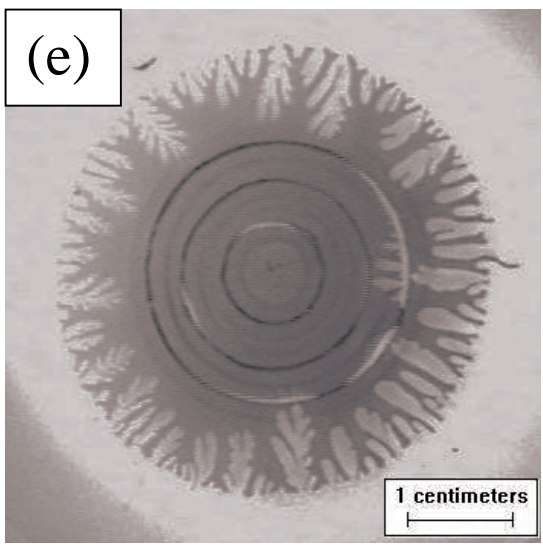
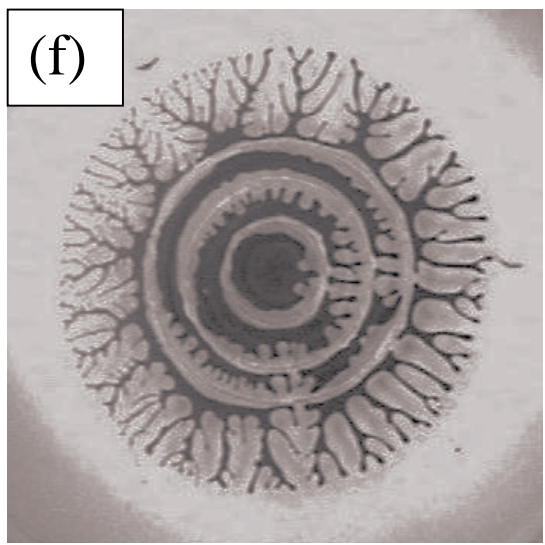
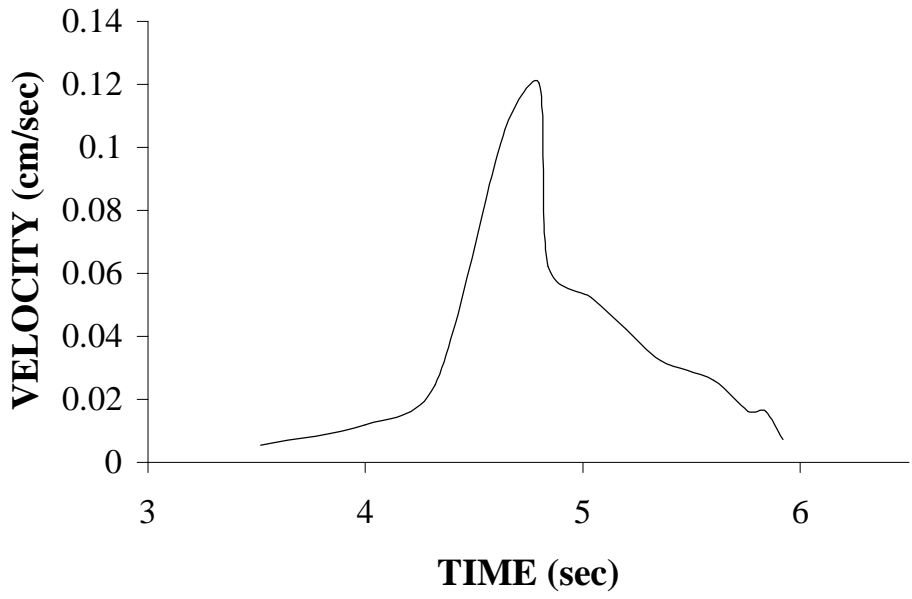
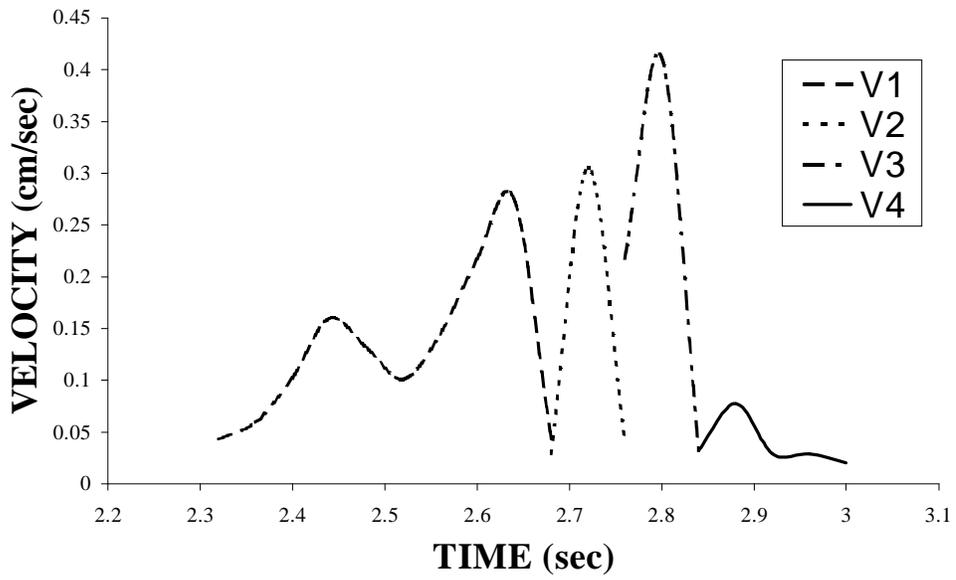
\end{document}